\documentclass[preprint,12pt]{elsarticle}

\usepackage{graphicx,pstricks}

\usepackage{amsmath,amssymb}

\journal{Physics Letters B}

\begin{document}

\begin{frontmatter}



\title{Charmonium suppression at RHIC and SPS: a hadronic baseline}




\author[wroclaw]{D.~Prorok}
\ead{prorok@ift.uni.wroc.pl}
\author[wroclaw]{L.~Turko}
\ead{lturko@ift.uni.wroc.pl}
\author[wroclaw,dubna]{D.~Blaschke}
\ead{blaschke@ift.uni.wroc.pl}

\address[wroclaw]{Institute for Theoretical Physics, University of
  Wroclaw, 50-204 Wroclaw, Poland}
\address[dubna]{Bogoliubov Lab.\ for Theoretical Physics, JINR Dubna,
  141980 Dubna, Russia}

\begin{abstract}
A kinetic equation approach is applied to model anomalous $J/\psi$
suppression at RHIC and SPS by absorption in a hadron resonance gas
which successfully describes statistical hadron production in both
experiments. The puzzling rapidity dependence of the PHENIX data is
reproduced as a geometric effect due to a longer absorption path for
$J/\psi$ production at forward rapidity.
\end{abstract}


\end{frontmatter}



\section{Introduction}
The effect of $J/\psi$  suppression  \cite{Matsui:1986dk} is one of
the diagnostic tools to prove quark-gluon plasma (QGP) formation in
heavy-ion collisions. After a rich phenomenology provided by the
CERN-SPS experiments NA3, NA38, NA50 and NA60, which nevertheless
could not unambiguously prove QGP formation, the situation became
even more complex due to first results from RHIC experiments PHENIX
and STAR. New experiments are planned at CERN-LHC and FAIR-CBM, for
a recent review, see \cite{Rapp:2008tf}. In order to make progress
in the interpretation of experimental findings it is important to
define benchmarks, such as a baseline from a statistical kinetic
model provided in this communication, which is based on a purely
hadronic description.

Measurements of $J/\psi$  suppression in Au-Au and Cu-Cu collisions
by STAR and PHENIX experiments
\cite{Adare:2006ns,Adare:2008sh,Abelev:2006db} brought up two
unexpected results. First, at forward rapidity  $J/\psi$ is  more
suppressed than at midrapidity. Second, the dependence of $J/\psi$
suppression at midrapidity on the number of participants $N_{\rm
part}$ \cite{Abelev:2006db, Adler:2004ta} coincides with the one
observed at SPS Pb-Pb collisions by the NA50 experiment
\cite{Ramello:2003ig}. Two different mechanisms to explain those
RHIC data are usually considered in the literature: recombination
\cite{Thews:2000rj,Grandchamp:2002wp,Capella:2007jv,Zhao:2008pp} or
statistical coalescence
\cite{BraunMunzinger:2000px,Andronic:2003zv,Kostyuk:2003kt,Andronic:2008gm}
and nuclear effects
\cite{Kharzeev:2008cv,Kharzeev:2008nw,Ferreiro:2008wc}.

We present a systematic step towards a general description of
$J/\psi$ absorption in the framework of a statistical analysis. The
formalism allows \cite{Prorok:2000kv,Prorok:2008zm} for a unified
description of the data from the NA38 and NA50 experiments at CERN
and from the PHENIX experiment at RHIC. Both PHENIX results,
concerning rapidity and $N_{\rm part}$ dependences of $J/\psi$
suppression in Au-Au and Cu-Cu collisions are simultaneously
reproduced.

The $J/\psi$ absorption is caused here by the effective hadronic
medium consisting of a multi-component non-interacting hadron
resonance gas (HRG). All hadrons from the lowest up to 2 GeV mass
are taken into account as constituents of the matter
\cite{Hagiwara:fs}. The gas is in thermal equilibrium and expands
longitudinally and transversally according to relativistic
hydrodynamics \cite{Baym:1983sr}. $J/\psi$ suppression is the result
of inelastic scattering on constituents of the HRG and on nucleons
in the cold nuclear matter (CNM) of the colliding ions. Both sources
of $J/\psi$ suppression, namely absorption in CNM and HRG, are
considered simultaneously.

The model applied here is a straightforward generalization of the
model of Refs.~\cite{Blaizot:1989ec,Prorok:2000kv} to include the
non-zero rapidity case. In our model, $J/\psi$ suppression is the
result of a $J/\psi$ final state absorption in a confined medium
through interactions of the type
\begin{equation}
J/\psi + h \longrightarrow D+ \bar D+X\\ ,  \label{psiabs}
\end{equation}
where $h$ denotes a hadron, $D$ is a charm meson and $X$ stands for
a particle which assure conservation laws (charge, baryon number,
strangeness). According to the above assumption, charmonium states
can be absorbed first in nuclear matter and soon after in the hadron
gas. In our phenomenological analysis we assume universal cross
sections for baryons, with the appropriate thresholds for their
dissociation reactions but energy-independent, $\sigma_{J/\psi N}=4$
mb. This choice corresponds to the absorption cross section
$\sigma_{\rm abs}=4.2 \pm 0.5$ mb for $J/\psi$ in cold nuclear
matter measured in p-A collisions by the NA50 experiment
\cite{Alessandro:2006jt}, shown to be compatible
\cite{Ferreiro:2008wc,Tram:2008zz} with recent d-Au data from the
PHENIX experiment \cite{Adare:2007gn}. The absorption cross section
for meson impact is taken as $\sigma_m=2 \sigma_{J/\psi N}/3$,
according to quark counting rules.

\begin{figure}[h]
\begin{center}
\resizebox*{!}{7.0cm}{\includegraphics{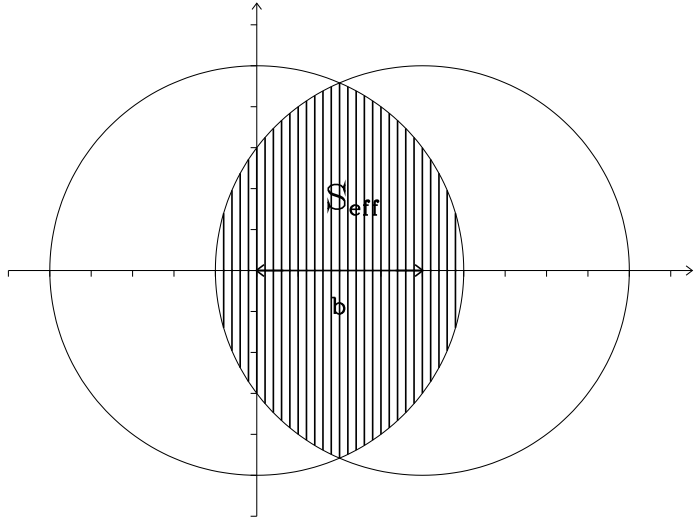}}
\end{center}
\caption{\label{Seff} View of a A-A collision at impact parameter
$b$ in the transverse plane $\vec{s}=(x,y,z=0)$. The region where
the nuclei overlap has been hatched and its area is $S_{eff}$.}
\end{figure}

An \mbox{A-A} collision at impact parameter $b$ generates an almond
shaped overlap region $S_{eff}$ as presented in Fig.~\ref{Seff}. The
time $t=0$ corresponds to the moment of the maximal overlap of the
nuclei. After about half of the time the nuclei need to cross each
other, matter appears in the central region. Soon thereafter matter
thermalizes and this moment, $t_{0}$, is estimated at about 1 fm. In
the real situation the longitudinal thickness of the matter at
$t_{0}$ is also of the order of 1 fm. Then matter starts to expand
and cool according to relativistic hydrodynamics (for details see
\cite{Baym:1983sr,Prorok:2000kv,Prorok:2008zm}) until reaching the
freeze-out temperature $T_{\rm f.o.}$ and subsequently freely
streaming towards the detectors.

A $c\bar{c}$ pair is created in a hard nucleon-nucleon collision
during the passage of the colliding nuclei through each other. It
will evolve to a charmonium eigenstate during the formation time
which is of the order of $t_{0}$ while experiencing cold nuclear
matter (CNM) effects. These are due to nuclear modification of the
initial state gluon distributions and partial extinction while
passing through the colliding nuclei and will be appropriately
parametrized using data from $pA$ collisions. We are interested here
to describe the ``anomalous suppression'' effect by subsequent final
state interactions with the expanding HRG formed between the
receding nuclei. Since these nuclei which border the HRG on both
sides in the longitudinal direction move almost with the speed of
light, even $J/\psi$s which contribute to the particle spectra
measured in the forward (backward) rapidity region can not escape
the hadronic medium. Due to the different production process, the
$J/\psi$ velocities can be considered as independent from the
velocity distribution of the HRG medium. This results in
considerable impact velocities, in particular for forward (backward)
produced $J/\psi$s traversing counter-propagating HRG flow.

The transverse expansion starts in the form of a rarefaction wave
moving inward $S_{eff}$. The evolution is decomposed into a
longitudinal expansion inside a slice bordered by the front of the
rarefaction wave and the transverse expansion. The rarefaction wave
moves radially inward with the sound velocity $c_{s}$. Since the
temperature (and hadron gas density) decreases rapidly outside the
wave, we shall ignore possible $J/\psi$ scattering there. We denote
by $t_{\rm esc}$ the moment of crossing the border of the
rarefaction front.

Inside the region bordered by the front of the rarefaction wave the
hydrodynamic evolution ceases when the freeze-out temperature is
reached, here we take $T_{\rm f.o.}=150$ MeV what is suggested by
the statistical model analysis of the PHENIX data
\cite{Rafelski:2004dp,Prorok:2007xp}. So $J/\psi$ is subject to
absorption for the time $t_{\rm final}={\rm min}\{ \langle t_{\rm
esc}\rangle,t_{\rm f.o.}\}$ where $\langle t_{\rm esc}\rangle$ is
the mean time until it escapes the hadron gas region and $t_{\rm
f.o.}$ is the time when it passes the kinetic freeze-out surface for
its absorption reactions. The rapidity dependence of $t_{\rm final}$
is crucial for the description of $J/\psi$ RHIC absorption
processes.

\begin{figure}[h]
\begin{center}
\resizebox*{!}{7.0cm}{\includegraphics{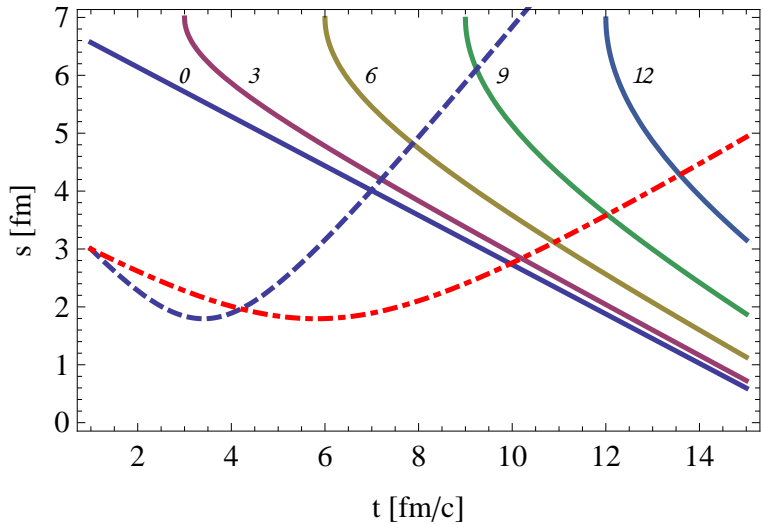}}
\end{center}
\caption{\label{rarefaction} $J/\psi$ (dashed and dot-dashed lines)
inside hadronic gas absorption region and the time evolution of the
rarefaction front radius $s$ at different $z$ (continuous lines with
labels denoting $z$ in fm). Numerical values correspond to  Au-Au
collisions.}
\end{figure}

For comparison with data one should estimate the number of
participating nucleons $N_{\rm part}(b)$. The Woods-Saxon nuclear
matter density distribution for Au and Cu nuclei is assumed with
parameters taken from \cite{De Jager:1974dg} and the standard
expression for $N_{\rm part}(b)$ is used
\cite{Bialas:1976ed,Prorok:2000kv}.

\section{$J/\psi$ absorption}

The $J/\psi$ absorption processes in CNM (nuclear absorption - NA)
and HRG take place subsequently. Due to this separation in time the
$J/\psi$ survival probability for given rapidity in a heavy-ion
collision with impact parameter $b$ assumes the form
\begin{equation}
R_{AA}(y,b) = S_{\rm NA}(y,b) \cdot S_{\rm HRG}(y,b)\ ,
\label{Surfact}
\end{equation}
where  $S_{\rm NA}(y,b)$ and $S_{\rm HRG}(y,b)$ are $J/\psi$
survival probabilities in CNM and HRG, respectively. For $S_{\rm
NA}$ we employ the usual approximation
\begin{equation}
S_{\rm NA} \cong \exp \left\{ -\sigma_{\psi N} \rho_0 \langle
L\rangle \right\}\ , \label{Survnmat}
\end{equation}
where $ \langle L \rangle$ is the mean absorption path length of the
$J/\psi$ through the colliding nuclei obtained from the Glauber
model and $\rho_0$ is the average nuclear matter density determined
from the normalization of the Woods-Saxon distribution to the mass
number $A$ for each nucleus separately. Within this approximate
expression, $S_{\rm NA}$ does not depend on rapidity.

The HRG survival probability $S_{\rm HRG}(y,b)$ is defined as
\begin{equation}
S_{\rm HRG}(y,b)= \frac{\int d^{2}p_{T}\int d^{3}r\;
\mathcal{F}(\vec{r},y,\vec{p}_{T},t)|_{t=\infty}}{\int
d^{2}p_{T}\int
d^{3}r\;\mathcal{F}(\vec{r},y,\vec{p}_{T},t)|_{t=t_0}}\; ,
\label{Survivfet}
\end{equation}
where the initial time $t_{0} = 1$ fm denotes the moment of the
thermalization of the created hadronic matter and the beginning of
the hydrodynamical expansion.

The rapidity-momentum $J/\psi$ distribution results from the
kinematical change of variables $d^3p\to d^2p_T dy$ and is given by
$\mathcal{F}(\vec{r},y,\vec{p}_{T},t) = M_{T} \cosh (y)$ $\times
f(\vec{r},\vec{p},t)$, where

\begin{eqnarray}
f(\vec{r},\vec{p},t) &=& f_{0}(\vec{r}-\vec{v}(t-t_{0}),\vec{p})
\\ & & \times \exp \left\{ -\int_{t_{0}}^{t} dt' \sum_{i=1}^{l} \int
{ {d^{3}q} \over {(2\pi)^{3}} } f_{i}(\vec{r}-\vec{v}(t-t'),
\vec{q},t') \sigma_{i} v_{rel}^{i} { {p_{\nu}q^{\nu}_{i}} \over
{EE^{\prime}_{i}} } \right\}\nonumber \label{Formlsol}
\end{eqnarray}
is the formal solution of a kinetic equation (for details see
\cite{Blaizot:1989ec,Prorok:2000kv}). The sum is over all considered
species of hadronic scatters with distributions
$f_{i}(\vec{r},\vec{q},t)$.

We have assumed that the initial distribution factorizes into
$\mathcal{F}(\vec{r},y,\vec{p}_{T},t_0)$ $= f_{0}(\vec{r})\cdot
g_{0}(p_{T})\cdot h_{0}(y)$ with $f_{0}(\vec{r})$ and $g_{0}(p_{T})$
normalized to unity. It is clear that $h_{0}(y)$ has not to be
specified since it cancels in the ratio (\ref{Survivfet}). Because
of the arguments given in the Introduction we shall assume that both
the created $J/\psi$ and the hadronic medium are included in $z=0$
plane. Thus the distribution of $J/\psi$ at $t_{0}$ is
$f_{0}(\vec{s})$ where $\vec{s}$ belongs to the $z=0$ plane. For
$g_{0}(p_{T})$ we take the form suggested by the initial state
interactions
\cite{Hufner:1988wz,Gavin:1988tw,Blaizot:1989hh,Gupta:1992cd}
\begin{equation}
 g_{0}(p_{T}) = { {2p_{T}}
\over {\langle p_{T}^{2}\rangle_{J/\psi}^{AA}} } \cdot \exp \left\{-
{{ p_{T}^{2}} \over {\langle p_{T}^{2}\rangle_{J/\psi}^{AA}} }
\right\}\ , \label{transv}
\end{equation}
where $\langle p_{T}^{2}\rangle_{J/\psi}^{AA}$ is the mean squared
transverse momentum of $J/\psi$ measured in an Au-Au (Cu-Cu)
collision \cite{Adare:2006ns,Adare:2008sh}.

For the description of the evolution of the matter, relativistic
hydrodynamics is employed. The longitudinal component of the
solution of the hydrodynamic equations (the exact analytic solution
for an (1+1)-dimensional case) reads \cite{Bjorken:1983qr}
\begin{equation}
s(\tau)= { {s_{0}\tau_{0}} \over \tau } \;,\;\;\;\;\;\;\;\;\;
n_{B}(\tau)= { {n_{B}^{0}\tau_{0}} \over \tau }\;, \label{bjork}
\end{equation}
%
where $\tau = \sqrt{t^{2}-z^{2}}$ is the proper time and $s_{0}$
($n_{B}^{0}$) is the initial density of entropy (baryon number).

For $n_{B}=0$ and a uniform initial temperature distribution with a
sharp edge at the border established by the nuclear surfaces, the
full solution of the (3+1)-dimensional hydrodynamic equations is
known \cite{Baym:1983sr}.

The hadron distributions have the usual form:
\begin{equation}
f_{i}(\vec{r},\vec{q},t)= { {2s_{i}+1} \over { \exp \left\{ {{
q_{i}^{\nu}u_{\nu}(\vec{r},t) - \mu_{i}(\vec{r},t) } \over
T(\vec{r},t)} \right\} + g_{i} } }\;, \label{Hadrdistr}
\end{equation}
where $u^{\nu}(\vec{r},t)$ is the four-vector the hadron flow
velocity and $m_{i}$, $B_{i}$, $S_{i}$, $\mu_{i}$, $s_{i}$ and
$g_{i}$ are the mass, baryon number, strangeness, chemical
potential, spin and a statistical factor of species $i$,
respectively. We treat an antiparticle as a different species and
introduce the chemical potential profile $\mu_{i}(\vec{r},t) =
B_{i}\mu_{B}(\vec{r},t) + S_{i}\mu_{S}(\vec{r},t)$, where
$\mu_{B}(\vec{r},t)$ and $\mu_{S}(\vec{r},t)$ are the local baryon
number and strangeness chemical potentials, respectively.

Since for the flow described in \cite{Baym:1983sr} the radial
velocity is zero inside the region bordered by the rarefaction wave
and $\textrm{v}_{z}(\vec{s},z,t) = z/t$, the four-velocity
$u^{\nu}(\vec{r},t)$ simplifies to
\begin{equation}
u^{\nu}(\vec{r},t) = {1 \over \tau} (t, 0, 0, z)\;.
\label{Fourveloc}
\end{equation}
Note that a $J/\psi$ with longitudinal velocity $v_{L}$ is at
$z=v_{L}(t-t_{0})$ at the moment $t$ and the longitudinal flow at
this point equals $\textrm{v}_{z}(\vec{s},z,t) = z/t =
v_{L}(t-t_{0})/t < v_{L}$, that is the relative longitudinal
velocity of $J/\psi$ with respect to the flow is always non-zero.
This reflects the fact that only the evolution of the medium is
boost-invariant but not of the $J/\psi$. The latter is moving along
a straight line with the constant velocity from the point of
creation as it does not take part in the thermalization of the
hadronic medium.

The temperature also does not depend on the radial coordinate $s$
within the region bordered by the rarefaction wave,
\begin{equation}
T(\vec{r},t) = T(z,t) = T(0,\tau) = T_{0} \cdot \tau^{-a}\;
\label{Tempfunct}
\end{equation}
and the power $a$ is the sound velocity squared at $T_{0}$,
$a=c_{s}^2(T_{0})$ \cite{Prorok:2000kv,Prorok:2008zm}.

To obtain $T_{0}$ and $a$ the following procedure is applied. For a
given centrality bin the initial energy density $\epsilon_{0}$ is
estimated in \cite{Adler:2004zn} (Au-Au) and
\cite{Rakotozafindrabe:2007zz} (Cu-Cu). We put $n_{S} = 0$ for the
strangeness number density and $n_{B}^{0}= 0$ for the initial baryon
number density. It should be noted here that the $J/\psi$
suppression pattern is practically the same for values of $n_{B}^{0}
\leq 0.65$ fm$^{-3}$ in this model as it was shown for the case of
the centrality dependence in Ref.~\cite{Prorok:2000kv}. This range
of $n_{B}^{0}$ corresponds to the possible values of $\mu_{B}$ up to
$\sim 300$ MeV during the whole evolution. This is far above the
estimated values of $\mu_{B} = 20-30$ MeV at the top RHIC energy
\cite{Prorok:2007xp,Andronic:2008gu}. Now, expressing
$\epsilon_{0}$, $n_{B}^{0}$ and $n_{S}$ as functions of $T$,
$\mu_{B}$ and $\mu_{S}$ in the grand canonical ensemble framework,
we can obtain $T_{0}$, $\mu_{B}^{0}$ and $\mu_{S}^{0}$. In our case
$\mu_{B} = \mu_{B}^{0} = 0$ and $\mu_{S} = \mu_{S}^{0} = 0$ always,
so the temperature is the only significant statistical parameter
here.

For the most central collisions we have obtained $T_{0} = 222.6$ MeV
($\epsilon_{0} = 5.4$ GeV/fm$^{3}$) for Au-Au and $T_{0} = 201.8$
MeV ($\epsilon_{0} = 2.5$ GeV/fm$^{3}$) for Cu-Cu. With $T_{0}$
known the initial entropy density $s_{0}$ is calculated. Having put
$s_{0}$ into (\ref{bjork}) and with the use of the grand canonical
expressions for $s$ we can obtain $T(\tau)$. It turned out that
$T(\tau)$ has the form given by (\ref{Tempfunct}) with $a$ varying
in the range $a = 0.148 - 0.156$ for $\epsilon_{0}$ in the range
$\epsilon_{0} = 5.5 - 0.5$ GeV/fm$^{3}$.

For a $J/\psi$ which is at $\vec{s}_{0} \in S_{eff}$ at the moment
$t_{0}$ and has the velocity $\vec{v} = \vec{v}_{T} + \vec{v}_{L} =
\vec{p}_{T}/E + \vec{p}_{L}/E$ we denote by $t_{\rm esc}$ the moment
of crossing the border of the rarefaction front.

The rarefaction front forms the ellipse in the $z,s$ plane
\cite{Baym:1983sr}
\begin{equation}
    (s-R_A)^2 + c_s^2 z^2 = c_s^2 t^2
\end{equation}
The continuous lines in Fig.~\ref{rarefaction}, labeled by different
values of the longitudinal variable $z$, show the time evolution of
the rarefaction front radius $s$ at the position $z$.

The $J/\psi$  trajectory is given by equations
\begin{eqnarray}
  \vec{s} &=& \vec{d} + \vec{v}_T (t-t_0)\,, \\
  z &=& v_L(t-t_0)\,.
\end{eqnarray}
We have used here the notation $\vec{d} = \vec{s}_{0} - \vec{b}$ for
the $J/\psi$ which escapes via the left half of $S_{eff}$ and
$\vec{d} = \vec{s}_{0}$ in the opposite case.

The dashed and dash-dotted lines in Fig.~\ref{rarefaction} are
$(s,t)$ trajectories of $J/\psi$ particles originating from the same
point in the region $S_{eff}$ shown in Fig.~1, but with different
values of transverse velocity $\vec{v}_T$. The dot-dashed line
corresponds here to the lower value of $v_T$.

The rarefaction front crossing time $t_{\rm esc}$ is a solution of
the equation

  \begin{equation}
\label{escape} \sqrt{d^2 + 2 \vec{d}\cdot\vec{v}_T (t-t_0) +
v_T^2(t-t_0)^2} = R_A - c_s\sqrt{t^2 - v_L^2(t-t_0)^2}\,.
\end{equation}

Having obtained $t_{\rm esc}$, we average it over the angle between
$\vec{s}_{0}$ and $\vec{v}_{T}$. Finally, to obtain $\langle t_{\rm
esc}\rangle(b,v_T,v_L)$ - the average time it takes the $J/\psi$ to
leave the hadronic medium, when it is produced in a collision at
impact parameter $b$ with the velocity $v_T=\sqrt{v_x^2+v_y^2}$ and
$v_L$ - we average the result over $S_{eff}$ with the weight given
by the initial $J/\psi$ distribution $f_{0}(\vec{s}_{0})$. This
weight is equal to
\begin{equation}\label{14}
f_{0}(\vec{s}_{0}) = \frac{T_{A}(\vec{s}_{0})T_{B}(\vec{s}_{0} -
\vec{b})}{T_{AB}(b)}
\end{equation}
where
\[T_{AB}(b) = \int~ d^2s~T_{A}(\vec{s}) T_{B}(\vec{s} - \vec{b})\,.\]
$T_{A}(\vec{s}) = \int\! dz\, \rho_{A}(\vec{s},z)$ is the nuclear
thickness function and $\rho_{A}(\vec{s},z)$ the Woods-Saxon nuclear
density distribution.

Then the final expression for $S_{\rm HRG}(y,b)$ reads

\begin{eqnarray}
S_{\rm HRG}(y,b) &=& {1 \over \int dp_{T}\; M_{T}\; g_{0}(p_{T})}
\int dp_{T}~M_{T}~g_{0}(p_{T})\label{Survynp} \\ & &\times\exp
\bigg\{ -\int_{t_{0}}^{t_{\rm final}} dt \sum_{i=1}^{l} \int
\frac{d^{3}q}{(2\pi)^{3}} f_{i}(\vec{q},t) \sigma_{i} v_{{\rm
rel},i}\frac{p_{\nu}q_{i}^{\nu}}{EE^{\prime}_{i}} \bigg\}\ ,
\nonumber
\end{eqnarray}
where $t_{\rm final}={\rm min}\{ \langle t_{\rm
esc}\rangle,t_{f.o.}\}$ and $t_{f.o.}$ is the freeze-out moment
resulting from the longitudinal Bjorken expansion.

Eq. (\ref{Survynp}) is in fact an approximation of the more involved
exact formula. The approximation means that the average of integrals
over trajectories has been replaced by the integration over the
averaged trajectory, similarly as in Eq. (\ref{Survnmat}). Our
preliminary results of estimations of the exact formula suggest that
the approximation influences only the normalization and in such a
way that this can be compensated by the rescaling of the absorption
cross-section to the lower values.

\section{Results}

\begin{figure}[!th]
\begin{center}
\resizebox*{!}{7.0cm}{\includegraphics{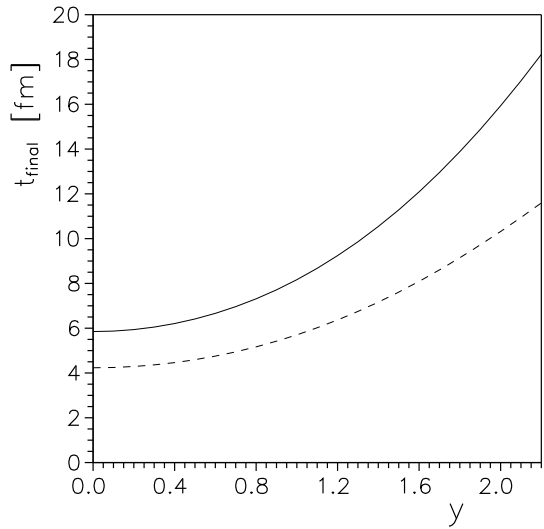}}
\end{center}
\caption{\label{Fig.2} An example of $t_{\rm final}$ in a function
of rapidity for $p_{T}= 2$ GeV and the $0-20 \%$ centrality bin for
Au-Au (solid) and Cu-Cu (dashed) collisions.}
\end{figure}

In Fig. \ref{Fig.2} an example of the behavior of $t_{\rm final}$
with rapidity is presented. This is for the case of the \mbox{$0-20
\%$} centrality bin of Au-Au and Cu-Cu collisions and for $p_{T}= 2$
GeV, taken as a typical value for collisions with a $\langle
p_{T}^{2}\rangle_{J/\psi} \sim 4$ GeV$^{2}$
\cite{Adare:2006ns,Adare:2008sh}.

\begin{figure}[!th]
\begin{center}
\resizebox*{!}{12.0cm}{\includegraphics{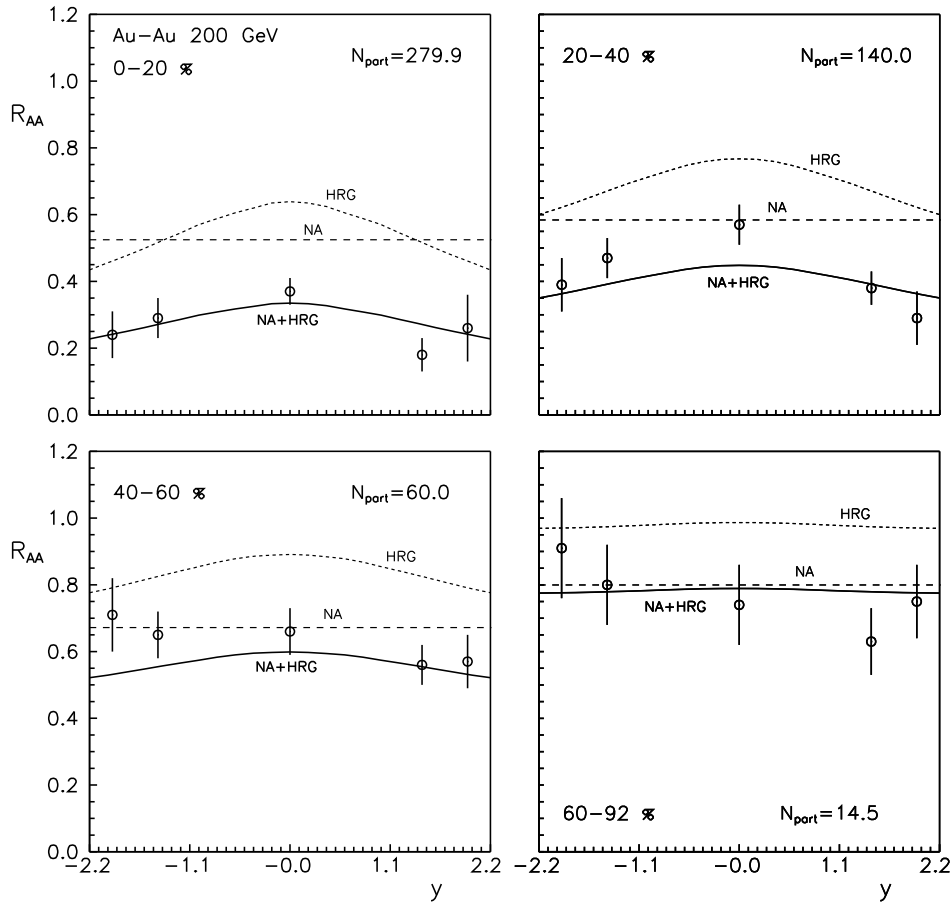}}
\end{center}
\caption{\label{Fig.3} $J/\psi$ nuclear modification factor versus
rapidity in Au-Au collisions for constant $\sigma_{J/\psi N}=4$ mb.
PHENIX data are from \cite{Adare:2006ns}. Errors shown are the
quadratic sum of statistical and uncorrelated systematic
uncertainties.}
\end{figure}

\begin{figure}[!th]
\begin{center}
\resizebox*{!}{12.0cm}{\includegraphics{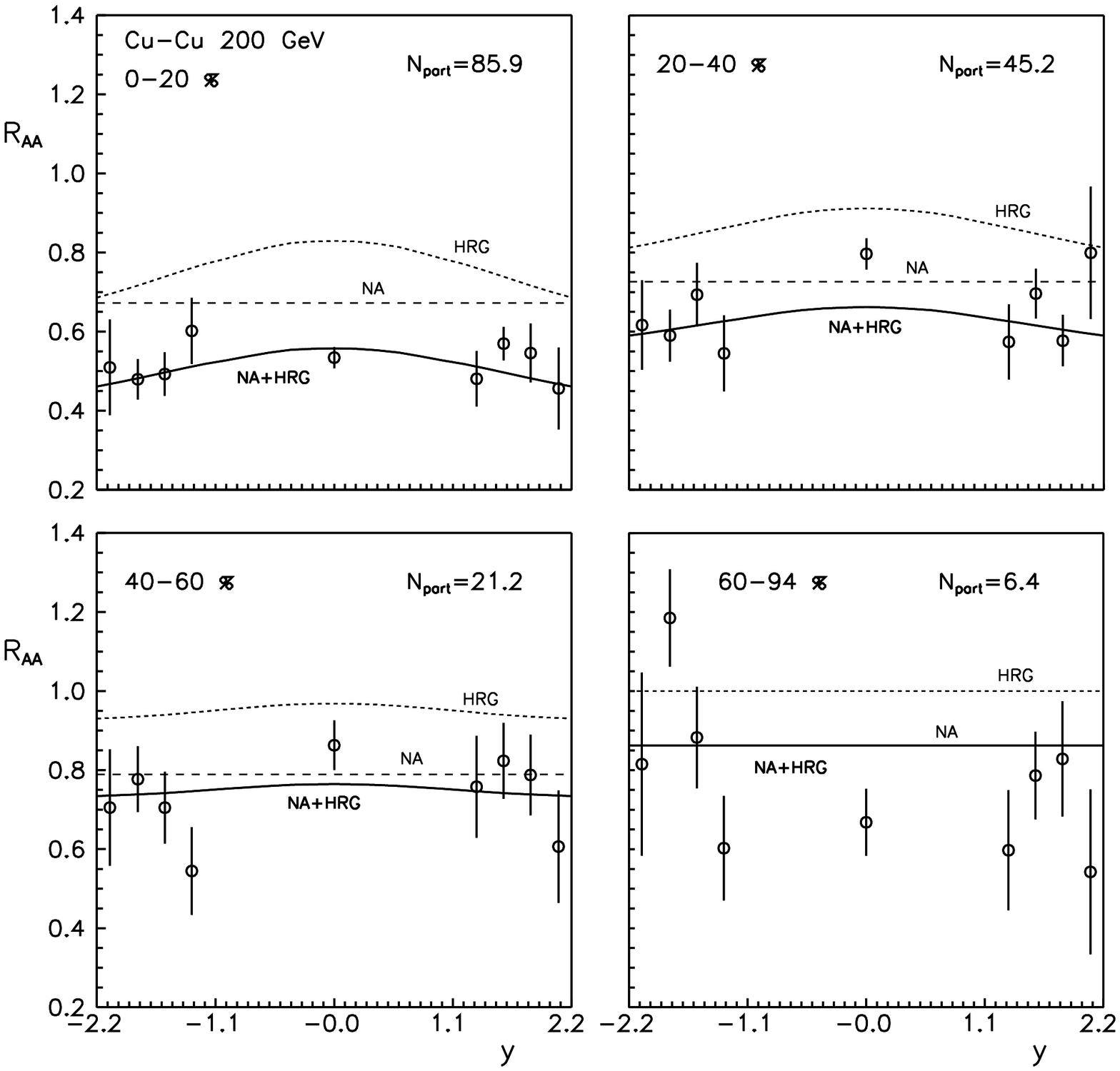}}
\end{center}
\caption{\label{Fig.4} Same as Fig. \protect\ref{Fig.3} but for
Cu-Cu collisions with PHENIX data from \cite{Adare:2008sh}. }
\end{figure}

In Figs. \ref{Fig.3} (\ref{Fig.4}) we show the $J/\psi$ nuclear
modification factor $R_{AA}$  for different centralities versus
rapidity in Au-Au (Cu-Cu) collisions vs. PHENIX data. There is an
overall agreement with the data.

\begin{figure}[!th]
\begin{center}
\resizebox*{!}{6.0cm}{\includegraphics{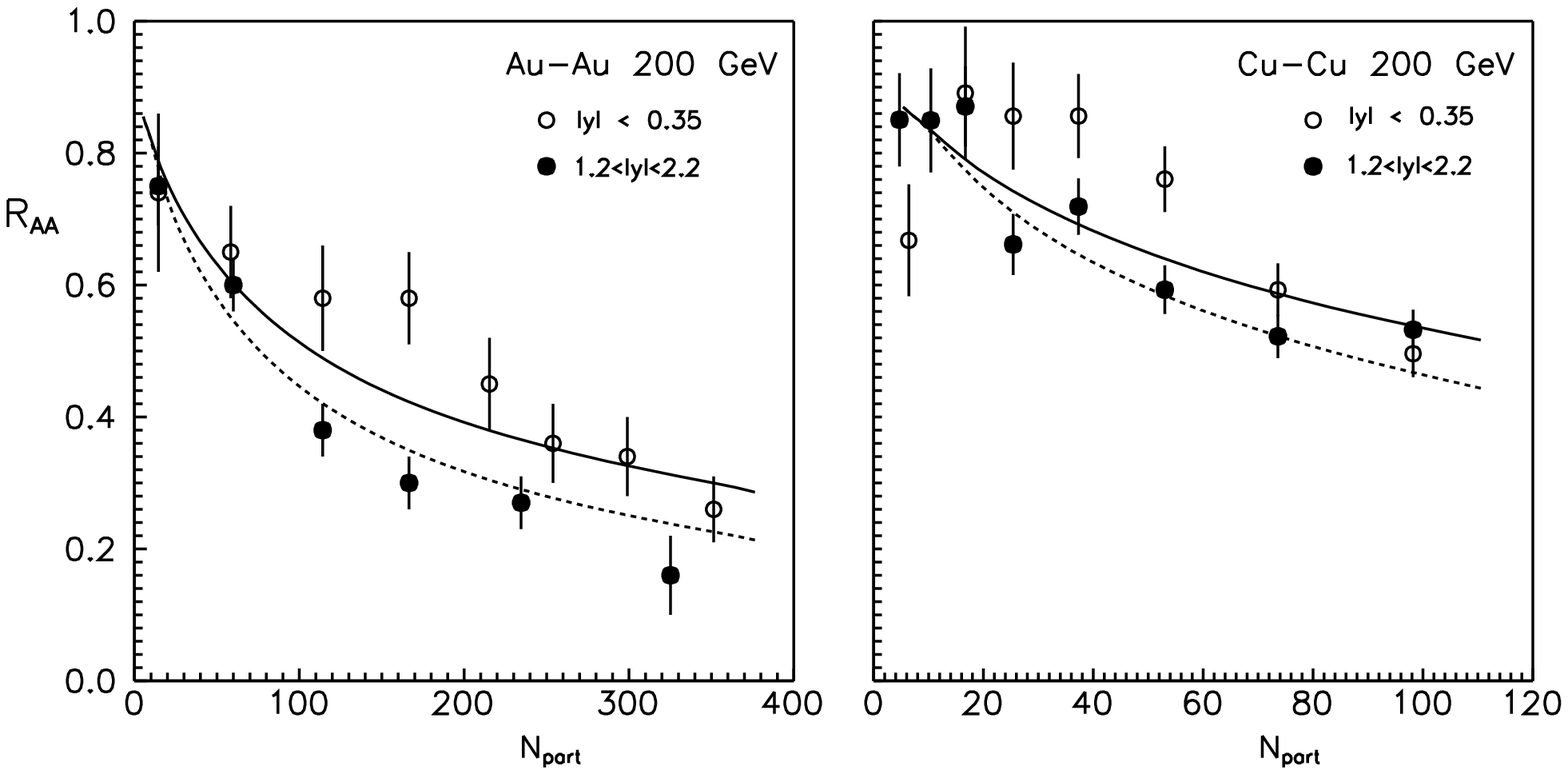}}
\end{center}
\caption{\label{Fig.5} $J/\psi$ nuclear modification factor versus
centrality in Au-Au (left) and Cu-Cu (right) collisions  for
constant $\sigma_{J/\psi N}=4$ mb. Lines are predictions of the
model for midrapidity (solid) and forward rapidity (dashed). PHENIX
data are from \cite{Adare:2006ns,Adare:2008sh}. Errors shown are the
quadratic sum of statistical and uncorrelated systematic
uncertainties. }
\end{figure}

The main new effect of the PHENIX data, that   $J/\psi$ suppression
is stronger at forward rapidity than at midrapidity is shown in Fig.
\ref{Fig.5} for the $N_{\rm part}$ dependence. It arises in the
present approach as a result of the geometry of the kinetic
freeze-out surface for the $J/\psi$ which extends much more in the
longitudinal direction for sufficiently long times, $t \geq R_{A} -
b/2$. This entails a longer absorption path for $J/\psi$ production
at forward rapidity.

Note that the present model simultaneously accounts for the
anomalous $J/\psi$ suppression in the centrality dependence of the
NA50/NA60 experiments at CERN, see
\cite{Prorok:2000kv,Prorok:2008zm}.
Can this result, obtained within a purely hadronic description, be
interpreted as evidence against QGP formation in these experiments?
Not at all, because of two reasons.

First, we have employed hadronic absorption cross sections which
cannot be reconciled with a microscopic description in, e.g.,
nonrelativistic \cite{Martins:1994hd,Barnes:2003dg} or relativistic
\cite{Ivanov:2003ge,Bourque:2008es} quark models which have a fastly
decreasing energy dependence and do not exceed a peak value of $2$
mb, see also \cite{Blaschke:2008mu}. In order to justify the
magnitude of hadronic absorption cross sections employed in the
present work, a strong medium dependence is required which could for
instance stem from spectral broadening of light mesons at the chiral
restoration transition (mesonic Mott effect) \cite{Burau:2000pn}.

Second, the hadronic resonance gas model gives a perfect decription
of the Lattice QCD thermodynamics in the vicinity of the critical
temperature $T_c$ for the chiral and deconfinement transition
\cite{Karsch:2003vd,Prorok:2008zm}. It has been demonstrated that
the description of QCD thermodynamics within a hadronic basis can be
extended even up to temperatures of $1.5~T_c$, provided a suitable
spectral broadening of these states above $T_c$ due to their Mott
effect is taken into account \cite{Blaschke:2003ut}. For a recent
development, see \cite{Bugaev:2008jj}. Such a Mott-Hagedorn
resonance gas model has been used to describe anomalous $J/\psi$
suppression at SPS \cite{Blaschke:2004xg} and is in accordance with
the picture of a strongly coupled QGP (sQGP) discovered at RHIC
\cite{Shuryak:2003xe,Lacey:2005qq}.

Summarizing this discussion, the present model may be seen as a
preparatory step towards a unified description of charmonium
suppression kinetics in SPS and RHIC experiments within a quantum
statistical approach to the sQGP to be developed. The geometrical
effect on the rapidity dependence decribed in the present work shall
be a part of such a description.

\section*{Acknowledgments}

We thank Susumu Oda for providing us with the $J/\psi$ rapidity data
for various centralities of Cu-Cu collisions and Krzysztof Redlich
for valuable discussions. This work was supported in part by the
Polish Ministry of Science and Higher Education under contract No. N
N202 0953 33.












\end{document}